\documentclass[conference]{IEEEtran}
\IEEEoverridecommandlockouts
\usepackage{cite}

\usepackage{amsmath,amssymb,amsfonts,mathtools,bm}
\usepackage{amsthm}
\usepackage{algorithm}
\usepackage{algorithmic}
\usepackage{graphicx}
\usepackage{textcomp}
\usepackage{xcolor,float}
\usepackage{tabularx}
\usepackage{textcomp}
\usepackage{diagbox}
\usepackage{svg}
\usepackage{booktabs}
\usepackage{subfigure}
\usepackage{hyperref}

\allowdisplaybreaks
\renewcommand{\baselinestretch}{1}

\begin{document}

\title{Constructing and Evaluating Digital Twins: An Intelligent Framework for DT Development}
\author{
\IEEEauthorblockN{
Longfei Ma\IEEEauthorrefmark{1},
Nan Cheng\IEEEauthorrefmark{1},
Xiucheng Wang\IEEEauthorrefmark{1},
Jiong Chen\IEEEauthorrefmark{1}\IEEEauthorrefmark{2},
Yinjun Gao\IEEEauthorrefmark{3},
Dongxiao Zhang\IEEEauthorrefmark{3},
Jun-Jie Zhang\IEEEauthorrefmark{3}\\
}
\IEEEauthorblockA{
\IEEEauthorrefmark{1}School of Telecommunications Engineering, Xidian University, Xi'an, 710071, China\\
\IEEEauthorrefmark{2}School of Telecommunications Engineering, Heriot-Watt University, Edinburgh, EH14 4AS, England\\
\IEEEauthorrefmark{3}Northwest Institute of Nuclear Technology\\
Email: \{lfma, xcwang\_1, jiongc\}@stu.xidian.edu.cn, dr.nan.cheng@ieee.org,\\ \{gaoyinjun, zhangdongxiao, zhangjunjie\}@nint.ac.cn,}}

    \maketitle

\IEEEdisplaynontitleabstractindextext

\IEEEpeerreviewmaketitle

\begin{abstract}
The development of Digital Twins (DTs) represents a transformative advance for simulating and optimizing complex systems in a controlled digital space. Despite their potential, the challenge of constructing DTs that accurately replicate and predict the dynamics of real-world systems remains substantial. This paper introduces an intelligent framework for the construction and evaluation of DTs, specifically designed to enhance the accuracy and utility of DTs in testing algorithmic performance. We propose a novel construction methodology that integrates deep learning-based policy gradient techniques to dynamically tune the DT parameters, ensuring high fidelity in the digital replication of physical systems. Moreover, the Mean STate Error (MSTE) is proposed as a robust metric for evaluating the performance of algorithms within these digital space. The efficacy of our framework is demonstrated through extensive simulations that show our DT not only accurately mirrors the physical reality but also provides a reliable platform for algorithm evaluation. This work lays a foundation for future research into DT technologies, highlighting pathways for both theoretical enhancements and practical implementations in various industries.
\end{abstract}


\section{Introduction}
As a revolutionary solution, digital twin (DT) technology has gained increasing attention in the field of wireless communications and networks \cite{8477101,9103025,9120192,9714139,9860495,9429703}. Compared to simulation platforms \cite{8994207}, DT focuses more on real-time with data, thus enabling better full-cycle mapping of physical entities. By accurately reproducing the characteristics of physical space in the digital realm, DT enables the testing of various algorithms and models to assess their impact on the physical space. This capability offers the potential to derive more effective strategies without escalating resource consumption or jeopardizing the integrity of physical systems. Consequently, a multitude of DT-based optimization and control strategies have emerged for wireless network optimization and industrial production. For instance, DT facilitates the analysis of specific behaviors' influence on wireless network transmission or industrial production reliability in the digital space. Moreover, it enables cost-effective and efficient reinforcement learning training that integrates well with existing work \cite{8672604}. Despite the proliferation of DT-assisted optimization and control algorithms, existing works have largely overlooked a fundamental challenge: how to construct a robust and efficient DT system capable of addressing this need.

Although some prior work has explored the construction of DT systems \cite{10201145,9213649,10000717}, the focus has primarily been on the fundamental function of real-time replication of physical space characteristics. These efforts have mainly concentrated on real-time data transmission and the synchronization of multiple data sources. However, during this period, the core function of DT, which involves analyzing the performance of algorithm models in the physical space based on their performance in the digital space, was not fully utilized. This aspect is of paramount importance as it not only reduces the cost of testing algorithm performance in the physical space but also mitigates the inherent risks associated with deploying unreliable or incomprehensible algorithm models directly in the physical space. To achieve this, it is mathematically imperative for the digital space to possess the same characteristic distribution transfer probability as the physical space \cite{cheng2024enhanced}. Specifically, the element state distribution of future time in the digital space and the strategic distribution of model generation must align with the distribution in the physical space. Noteworthy, we do not consider the strong DT system here, where the predicted state in the digital space corresponds exactly to the real state in the physical space at a future time, as this would fall into the dilemma of Laplace determinism. This alignment allows the data distribution in the digital space to be utilized for algorithm model improvement or selection, enabling the deployment of models with optimal performance or minimal system risk in the physical space.
\begin{figure*}
    \centering
    \includegraphics[width=1\linewidth]{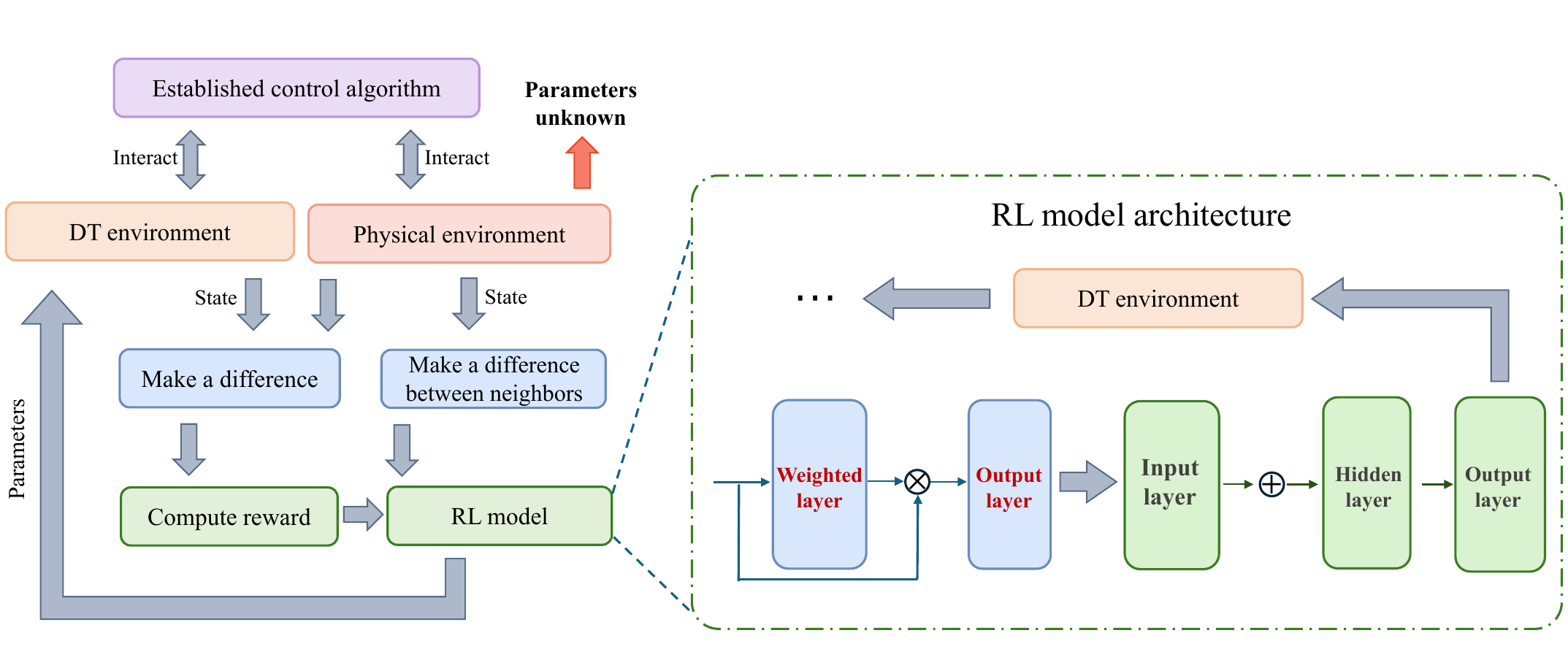}
    \caption{Illustration of proposed method.}
    \label{fig-system}
\end{figure*}

Therefore, it is imperative to reconsider how to construct a DT with the capability to assess algorithm model performance in the digital space. However, this task presents two significant challenges: the absence of recognized evaluation indicators and the lack of effective construction schemes. Unlike traditional test systems that aim to differentiate the performance of various algorithm models, the primary objective of a DT is to evaluate the impact of algorithm models on the physical space within the digital realm. In light of this, we propose a straightforward yet efficient evaluation metric called the mean state error (MSTE). The MSTE measures the disparity between the state sequences of the same algorithm after running for a specific duration in the physical space and the corresponding state sequences after running for the same duration in the digital space. A lower MSTE indicates a higher level of alignment between the performance of the digital space and the physical space running the same algorithm. The construction of a DT poses additional challenges, particularly when certain parameters in the physical space cannot be directly measured, and when the availability of observation data, such as physical space videos, is limited due to high physical interaction costs or other factors. However, it is precisely these challenges that underscore the importance of constructing a DT capable of faithfully reproducing the properties of the physical space in the digital domain. To address these challenges and achieve efficient DT construction, we propose an intelligent construction method based on the strategy layer. This approach enables the rapid derivation of state transition probability distribution parameters in the physical space. In summary, the main contributions of this paper are as follows. 
\begin{enumerate}
    \item To our best knowledge, we first study the problem of how to build a DT with the capability to test algorithm models in the digital space to obtain their performance in the physical space, and an MSTE metric is proposed to evaluate the test accuracy of the DT.
    \item A policy gradient-based is proposed to build the DT system, which allows for the rapid derivation of state transition probability distribution parameters in the physical space. 
\end{enumerate}

\section{System Model and Problem Formulation}
In this paper, we examine a scenario involving a physical space where the state transition probability distribution is unknown. Direct measurement cannot yield the relevant parameters of this physical space; however, observing the state changes through limited interactions is feasible. This setup can be generalized to other scenarios, such as infinite interactions with the physical space or scenarios where only limited-length videos are available, showcasing specific algorithm interactions within the physical space. Our objective is to construct a DT system that can accurately test the performance of algorithm models in the digital space, ensuring that this performance aligns closely with that in the physical space. We consider a physical system $F_p(\bm{s}_t, \bm{a}_t; \hat{\bm{\theta}})$, where the probability distribution is known but the distribution parameters $\hat{\bm{\theta}}$ are not, where $\bm{s}_t$ represents the state at time $t$, $\bm{a}_t$ denotes the input action at time $t$, and $\hat{\bm{\theta}}$ is the unknown distribution parameter. In this context, $\bm{\theta}$ can represent either a coefficient of a polynomial function or a parameter of the probability distribution. This paper focuses on the coefficient of the polynomial function, enabling function fitting through methods such as the Taylor series or Fourier series. Additionally, we explore a white-box testing scenario where the input $a_t$ of the physical system is generated by a known strategy $G(s_t)$. Due to the constraints imposed by the physical space, we observe that policy $G(\cdot)$ interacts with the physical space to produce an $N$-long sequence $\bm{\hat{s}} = [\hat{s}_1, \hat{s}_2, \cdots, \hat{s}_N]$, where $\hat{s}_i$ is the observed state in the physical space after the $i$-th interaction. Similarly, in a digital space with adjustable parameters $F_d(\cdot, \cdot; \bm{\theta})$, the strategy $G(\cdot)$ can interact with the digital space to yield an $N$-long sequence $\bm{s} = [s_1, s_2, \cdots, s_N]$, where $s_i$ is the state observed in the digital space after the $i$-th interaction with policy $G(\cdot)$. This formulation sets the stage for developing a robust DT system, as discussed in the following sections.

Although no standardized metric currently exists for assessing the suitability of a digital space for model or algorithm performance testing, considerable research addresses the challenge of sim2real performance degradation. This phenomenon arises when migrating models trained in simulated spaces to real-world settings. A prominent theory highlighting this dilemma is the concept of distribution shift between the simulation and physical domains. The extent of this shift directly correlates with the performance degradation experienced by models trained in simulation spaces when tested in real-world scenarios. This degradation occurs because models typically achieve consistent performance across identical distributed data. Thus, measuring the degree of distribution shift serves as a critical indicator of a digital space's capability to accurately replicate physical space characteristics. Specifically, the difference in input data distribution encountered by the same strategy $G(\cdot)$ in different spaces reflects the average discrepancy in the resulting state sequences. Consequently, we introduce the Mean State Transition Error (MSTE) as a metric for assessing whether the data in the digital space effectively mirrors the performance of models in the physical space, which is defined as follows.
\begin{align}
    \text{MSTE} = \sum_{i=1}^{N} \|\hat{\bm{s}}_{i}-\bm{s}_{i}\|,
\end{align}
Therefore, the problem of building a DT that can test the performance of an algorithmic model in digital space in physical space can be formulated as follows.
\begin{align}
    &\min_{\bm{\theta}}\, \sum_{i=1}^{N} \|\hat{\bm{s}}_{i}-\bm{s}_{i}\|,\label{obj}\\
    &s.t.\;\;\hat{\bm{s}}_{i+1}=F_{p}\left(\hat{\bm{s}}_{i},\hat{\bm{a}}_{i};\bm{\hat{\theta}}\right),\tag{\ref{obj}a}\\
    &\quad\;\;\;\;\hat{\bm{a}}_{i}=G(\hat{\bm{s}}_{i}),\tag{\ref{obj}b}\\
    &\quad\;\;\;\;\bm{s}_{i+1}=F_{d}\left(\bm{s}_{i},\bm{a}_{i};\bm{\theta}\right),\tag{\ref{obj}c}\\
    &\quad\;\;\;\;\bm{a}_{i}=G(\bm{s}_{i}),\tag{\ref{obj}d}
\end{align}

\section{Policy Gradient Based DT Construction}
\subsection{Feature Preprocessing Module}
The complexity of extracting meaningful features from raw data poses a significant challenge in various analytical tasks. To address this challenge effectively, a feature preprocessing module is indispensable \cite{8559182,10391672}. In our context, where we are confronted with the task of deducing distribution parameters, the initial preprocessing of features becomes even more critical. Unlike traditional approaches such as principal component analysis (PCA) \cite{1102568}, which may overlook the intricacies of the distribution parameter deduction problem, our paper advocates for a bespoke data preprocessing method tailored to this specific challenge. By systematically analyzing the unique characteristics of the problem at hand, we aim to develop a preprocessing technique that streamlines feature extraction, enabling more accurate and efficient analysis.

The intrinsic motivation guiding our proposed preprocessing method originates from a deep understanding of the relationship between distribution parameters and state transitions within physical and digital spaces. We posit that these parameters intricately govern the dynamics of system behavior, dictating the transitions between states when subjected to specific inputs from algorithm models. Recognizing the pivotal role of distribution parameters, our methodology prioritizes the preservation of the coupling relationship between adjacent data points. Traditional feature engineering dimensionality reduction techniques risk disrupting this relationship, undermining subsequent algorithmic analysis. Consequently, our approach advocates for a simple yet effective data preprocessing strategy: computing the difference between adjacent states to derive the sequence $\Delta\bm{\hat{s}}$, where $\Delta \hat{s}_{i}=\hat{s}_{i+1}-\hat{s}_{i}$. This transformed data serves as the cornerstone for subsequent DT constructions, facilitating more robust and insightful analyses of system dynamics and parameter influences.
\subsection{Policy Gradient Based Method}
To achieve efficient and high-performance DT construction, we employ a policy gradient (PG) \cite{sutton1999policy} method. This approach takes the physical space state sequence $\bm{\hat{s}}$, preprocessed through feature preprocessing, as input and outputs the estimated distribution parameters for the physical space. As depicted in Figure 1, these estimated physical space distribution parameters are then transferred to the digital space. Subsequently, the same strategy $G(\cdot)$ interacts with the environment in the digital space, yielding the digital space state sequence $\bm{s}$. To optimize the policy $G(\cdot)$ in the digital space and minimize the discrepancy between its performance in the digital and physical spaces $\|\bm{\hat{s}} - \bm{s}\|$, we employ a training method based on PG. Assuming, the PG is a neural network (NN) $\pi(\Delta \bm{\hat{s}};\bm{\varphi})$ with trainable parameters $\bm{\varphi}$, by defining $\Bar{\vartheta}=\int\vartheta_{\bm{\theta}}P(\bm{\theta})d\bm{\theta}=\int\sum_{i=1}^{N}\|\bm{\hat{s}}_i-\bm{s}_i\|P(\bm{\theta})d\bm{\theta}$ as the expectation of MSTE under different $\bm{\theta}=\pi(\Delta \bm{\hat{s}};\bm{\varphi})$, where $P(\bm{\theta})$ is the probability of the output of $\pi(\cdot)$ is $\theta$ which is written as $\pi(\Delta\bm{\hat{\theta}};\bm{\varphi})$. The gradient of $\bm{\varphi}$ can be derived as follows.
\begin{align}
    \nabla_{\bm{\varphi}}\Bar{\vartheta}=\int\vartheta_{\bm{\theta}}\nabla_{\bm{\varphi}}\pi(\Delta \bm{\hat{s}};\bm{\varphi})d\bm{\theta},\label{raw-grad}
\end{align}
Obviously, directly calculate the equation \eqref{raw-grad} is challenging due to the integration, however, we can simplify the equation \eqref{raw-grad} as follows.
\begin{align}
    \nabla_{\bm{\varphi}}\Bar{\vartheta}&=\int\vartheta_{\bm{\theta}}\pi(\Delta\hat{\bm{s}};\bm{\varphi})\frac{\nabla_{\bm{\varphi}}\pi(\Delta\bm{\hat{s}};\bm{\varphi})}{\pi(\Delta\bm{\hat{s}};\bm{\varphi})},\notag\\
    &=\int\vartheta_{\bm{\theta}}\pi(\Delta\hat{\bm{s}};\varphi)\nabla_{\bm{\varphi}}\log\pi(\Delta\bm{\hat{s}};\bm{\varphi}),\notag\\
    &=\mathbb{E}_{\bm{\theta}\sim\pi}[\vartheta_{\bm{\theta}}\nabla_{\bm{\varphi}}\log\pi(\Delta\hat{\bm{s}};\bm{\varphi})],\notag\\
    &\approx\sum \vartheta_{\bm{\theta}}\nabla_{\bm{\varphi}}\log\pi(\Delta\hat{\bm{s}};\bm{\varphi})\label{grad}.
\end{align}
According to equation \eqref{grad}, the parameters of $\bm{\varphi}$ can be updated by repeat the experiment independently, until the $\bm{\varphi}$ is updated good enough to find the optimal $\bm{\theta}$ to minimize the MSTE.

By organically combining the above two modules, the PG with feature preprocessing (PGFP) method proposed in this paper is obtained, and its overall flow is shown in Algorithm 1.

\begin{algorithm}[!ht]
    \small
	\caption{PGFP} 
	\begin{algorithmic}[1]
    \FOR{iteration = 1,$\ldots$, $\text{Iter}_{\operatorname{max}}$}
    \STATE Interacting with the physical environment using established strategy and collect state sequences
    \STATE Make a difference between neighboring states of physical environment as inputs
    \STATE Construct DT environment based on NN model outputs
    \STATE Interacting with the DT environment using established strategy and collect state sequences
    \STATE Compute rewards based on state information for both the DT environment and the physical environment
    \STATE Update the NN parameters based on the PG method as follows\\
    $\bm{\varphi} \gets \bm{\varphi} + \alpha\sum \vartheta_{\bm{\theta}}\nabla_{\bm{\varphi}}\log\pi(\Delta\hat{\bm{s}};\bm{\varphi})$
     \ENDFOR\\
    \RETURN the trained NN model $\pi(\Delta \bm{\hat{s}};\bm{\varphi})$

	\end{algorithmic}

\end{algorithm}

\section{Simulation Results}

In this study, we assess the efficacy of our proposed algorithm using a widely recognized environment for testing dynamic programming and reinforcement learning: the tower defense game environment. Within this environment, the original setup, characterized by specific parameters, constitutes the physical space. This space encompasses varying numbers of stochastic bosses and defense towers. The bosses possess freedom of movement within a 3D space, with a predefined objective of reaching a designated destination. Conversely, the role of defense towers is to thwart the advancement of bosses by engaging in attacks. Due to the towers' inability to modify their configuration, they must adjust their attack angles—namely, the horizontal and pitch angles—to effectively engage with the bosses. Given that the number of defense towers is typically fewer than the number of bosses, we employ the Hungarian algorithm as a known strategy $G(\cdot)$ to allocate attack targets for the defense towers. Subsequently, the defense towers adjust their attack angles based on the assigned targets to engage the bosses effectively. Notably, we consider the rotation angular velocity, a fundamental attribute of defense towers within the tower defense game, as an unknown parameter $\bm{\theta}$ requiring deduction. This elucidates the intricate interplay between game dynamics and algorithmic optimization strategies, forming the foundation for our investigative endeavor.

In the following results, in addition to the proposed PGFP, we set up three benchmark algorithms, described as follows.
\begin{itemize}
    \item $GAFP$: Compared with the proposed method, the PG part is replaced with the classical genetic algorithm (GA). However, the same feature preprocessing module is used and the reward is taken as the fitness. In this case, the population size is 30, the mutation rate is 0.01, the crossover rate is 0.7, and the selection rate is 0.5.
    \item $PG$: The original policy gradient method without using the feature preprocessing module. Directly use the state sequences as model inputs and construct the reward function using the MSTE of the DT environment and the physical environment.
    \item $GA$: The original GA without using the feature preprocessing module. 
\end{itemize}

\begin{figure}[ht]
  \centering
  \includegraphics[width=1.0\columnwidth]{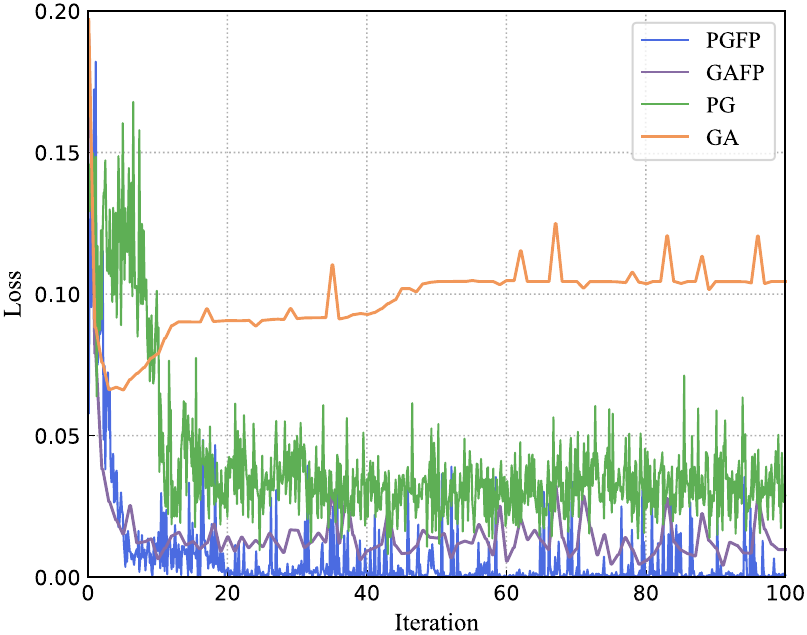}
   \vspace{-5pt}
  \centering \caption{Convergence performance of different methods.} 
  \label{loss}
   \vspace{-5pt}
\end{figure}

Fig. \ref{loss} shows the convergence performance of various methods under consideration. In an attempt to maintain consistency across the methods evaluated, we designate the x-axis to represent the iteration count. It should be noted that, within the methods related to PG, iteration count refers to the times with which the parameters within the NN model are updated. Conversely, in the case of methods associated with GA, iteration count signifies the number of times the population undergoes reproduction. The metric used to measure the training performance across all methods is the mean squared error (MSE) between the predicted parameter values and the actual, true parameter values. The figure demonstrates that the PGFP method outperforms the established benchmark methods by a significant margin. Furthermore, it is evident from the data that the inclusion of a feature preprocessing module markedly enhances the performance of the methods, as compared to their counterparts that lack such a module. Notably, despite the observed variations in effectiveness, all methods show a tendency to converge rapidly. This phenomenon may suggest a distinctive trait inherent to tasks that involving the deduction of environmental parameters: the primary challenge appears to stem more from accurately characterizing the features rather than the iterative refinement provided by the optimization algorithms.

\begin{figure}[ht]
  \centering
  \includegraphics[width=1.0\columnwidth]{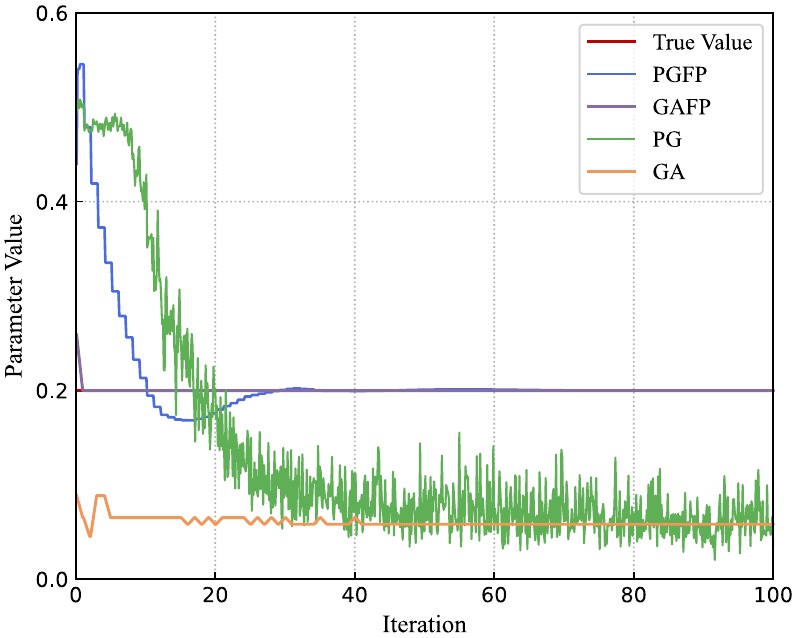}
   \vspace{-5pt}
  \centering \caption{Prediction of parameter 1 by different methods.} 
  \label{speed1}
   \vspace{-5pt}
\end{figure}

\begin{figure}[ht]
  \centering
  \includegraphics[width=1.0\columnwidth]{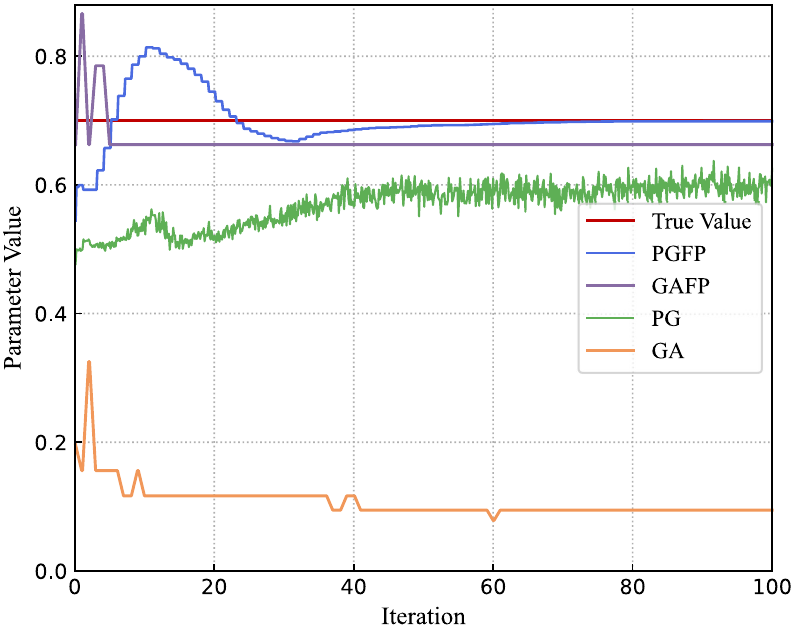}   \vspace{-5pt}  \centering \caption{Prediction of parameter 2 by different methods.} 
  \label{speed2}
   \vspace{-5pt}
\end{figure}

Fig. \ref{speed1} and Fig. \ref{speed2} demonstrate the comparative efficacy of different methods in predicting parameter values. The crimson delineations in the figures indicate the true values of the parameters, specifically 0.2 and 0.7 correspondingly. It can be seen that PGFP achieves extremely high prediction accuracy, and GAFP has similarly high accuracy in predicting parameter 1, while there is some deviation in predicting parameter 2. On the contrary, the original PG and GA algorithms exhibit poor performance, with PG marginally achieving minimally acceptable precision only in its prediction for parameter 2. This observation coheres with the outcomes delineated in Fig. \ref{loss}. Furthermore, it is pertinent to underscore that the original PG method exhibits substantial oscillations, a feature notably ameliorated upon the integration of the feature preprocessing module, implicating a notable enhancement in stability.

\section{Conclusion}
In this study, we have introduced a robust DT system capable of evaluating algorithm model performance in the digital space, exemplified by the MSTE metric, and an intelligent construction method based on the strategy layer for rapid derivation of state transition probability distribution parameters in physical spaces. The application of our findings extends to optimizing wireless networks and industrial production systems, offering opportunities to enhance network optimization and transmission reliability in wireless communications, as well as streamline production processes and improve system performance in industrial settings.

\bibliography{ref}
\bibliographystyle{IEEEtran}

\ifCLASSOPTIONcaptionsoff
  \newpage
\fi

\end{document}